\documentclass[prl,twocolumn,showpacs,floats]{revtex4-2}
\makeatletter
\def\maketitle{
\@author@finish
\title@column\titleblock@produce
\suppressfloats[t]}
\makeatother

\usepackage{latexsym}
\usepackage{amsmath}
\usepackage{amssymb}
\usepackage{bm}
\usepackage{wasysym}
\usepackage[dvips]{color}
\usepackage{graphicx}
\usepackage{subfigure}
\usepackage{epsfig}
\usepackage{pgfplots}
\usepackage{balance}
\usepackage{subfigure}
\usepackage{float}
\usepackage{comment}

\usepackage{natbib}
\usepackage{multibib}

\usepackage[normalem]{ulem}
\newcommand\redout{\bgroup\markoverwith
{\textcolor{red}{\rule[0.5ex]{2pt}{0.8pt}}}\ULon}

\DeclareRobustCommand{\vec}[1]{{%
    \bm{\mathbf{#1}}%
}}

\begin{document}

\title{Atoms in a spin dependent optical potential: ground state topology and magnetization}

\author{Piotr Szulim,$\,^1$ Marek Trippenbach,$\,^1$ Y. B. Band,$\,^2$ Mariusz Gajda,$\,^3$ and Miros{\l}aw Brewczyk$\,^4$}

\affiliation{
\mbox{$^1$ Faculty of Physics, University of Warsaw, ul. Pasteura 5, 02-093 Warszawa, Poland}
\mbox{$^2$ Department of Chemistry, Department of Physics,}
\mbox{Department of Electro-Optics, and the Ilse Katz Center for Nano-Science,} \\
\mbox{Ben-Gurion University, Beer-Sheva 84105, Israel} \\
\mbox{$^3$ Instytut Fizyki PAN, Aleja Lotnikow 32/46, 02-668 Warsaw, Poland}
\mbox{$^4$ Wydzia{\l} Fizyki, Uniwersytet w Bia{\l}ymstoku,  ul. K. Cio{\l}kowskiego 1L, 15-245 Bia{\l}ystok, Poland} }

\date{\today}
\begin{abstract}
We investigate a Bose-Einstein condensate of $F= 1$ $^{87}$Rb atoms in a 2D spin-dependent optical lattice generated by intersecting laser beams with a superposition of polarizations. For $^{87}$Rb the effective interaction of an atom with the electromagnetic field contains a scalar and a vector (called as fictitious magnetic field, $B_{fic}$) potentials. The Rb atoms behave as a quantum rotor (QR) with angular momentum given by the sum of the atomic rotational motion angular momentum and the hyperfine spin. The ground state of the QR is affected upon applying an external magnetic field, $B_{ext}$, perpendicular to the plane of QR motion and a sudden change of its topology occurs as the ratio $B_{ext}/B_{fic}$ exceeds critical value. It is shown that the change of topology of the QR ground state is a result of combined action of Zeeman and Einstein-de Haas effects. The first transfers atoms to the largest hyperfine component to polarize the sample along the field as the external magnetic field is increased. The second sweeps spin to rotational angular momentum, modifying the kinetic energy of the atoms.
\end{abstract}


\maketitle

\section{Introduction}
\label{sec:intro}

Ever since the experimental realization of Bose-Einstein condensates (BECs), they have proven to be an ideal platform for study of various effects of quantum many-body interaction. One of the milestones in experimental studies of BEC was the application of all optical traps \cite{aotrap} which liberated atom spin degrees of freedom from the trapping mechanism and allowed for much richer internal structures and dynamics of the BEC. The study of spinor BECs involves the interplay of spin-spin interactions, magnetization, topology, symmetry, homotopy, spin-gauge coupling, and magnetic solitons, etc., making it an incredibly fertile and promising research area \cite{Ho, Ueda1, Ueda3, Stamper, Aftalion, Stoof, Anderson, Fetter2, Chai}. For these studies important characteristics are spin textures and vortex spin textures in spinor BECs \cite{spin_textures}, which are related to superfluidity (see Ref.~\cite{superfluidity} for references) and topological states in condensed matter physics \cite{Lewy}. 
In our study we investigate bosonic cold atoms subject to a 2D spin-dependent optical lattice potential (SDOLP) in the context of spin textures and transition of the ground state topology (i.e., a change of symmetry of the many-body state upon varying an external parameter).  First we focus on the limit of singly occupied lattice sites and then on the limit of many atoms per lattice site, but the atoms are condensed such that the tunneling rate out of the sites is negligible (the so-called insulating regime). The superfluid phase, when tunneling between lattices sites are involved is also very exciting and will be the subject of future work.
Our work is closely related to the Einstein-de Haas (EdH) effect in BECs \cite{EdH, EdHUeda, EDHbis, EdHSwislocki}, where transfer of atoms to other Zeeman states can be observed due to dipolar interactions, which couple the spin and the orbital degrees of freedom.

The concept of a ``ficticious magnetic field'' for describing the vector component of the tensor polarizibility was first introduced by Cohen-Tannoudji in Ref.~\cite{Cohen}. It was later applied for quantum state control in optical lattices by Deutsch and Jessen \cite{Deutsch1, Deutsch2} and for dispersive quantum measurement of multilevel atoms \cite{Geremia}. It also led to engineering novel optical lattices \cite{Windpassinger}. For studies of alkali atoms in spin-dependent optical lattices created by elliptically polarized light beams see Ref.~\cite{Lekien}.  Recently a study of fermionic atoms in a 2D SDOLP in the limit of singly occupied sites was carried out in Ref.~\cite{Kuzmenko_19}. In addition to determining the wave functions and energy levels of the atoms in the SDOLP, it was shown that such systems could be used as high precision rotation sensors, accelerometers, and magnetometers.  Here we consider bosonic atoms and focus on the case of a BEC in the insulating regime.  Explicitly, we consider a Bose-Einstein condensate consisting of $^{87}$Rb atoms in a spin-dependent optical lattice potential SDOLP. We show that by applying an external magnetic field, the properties of the system are significantly enriched and a radical change of the topological properties of the ground state can be observed as the strength of the external field is varied.

\section{Model}
\label{sec:Model}

\begin{figure}
\centering
{\includegraphics[width=\linewidth]{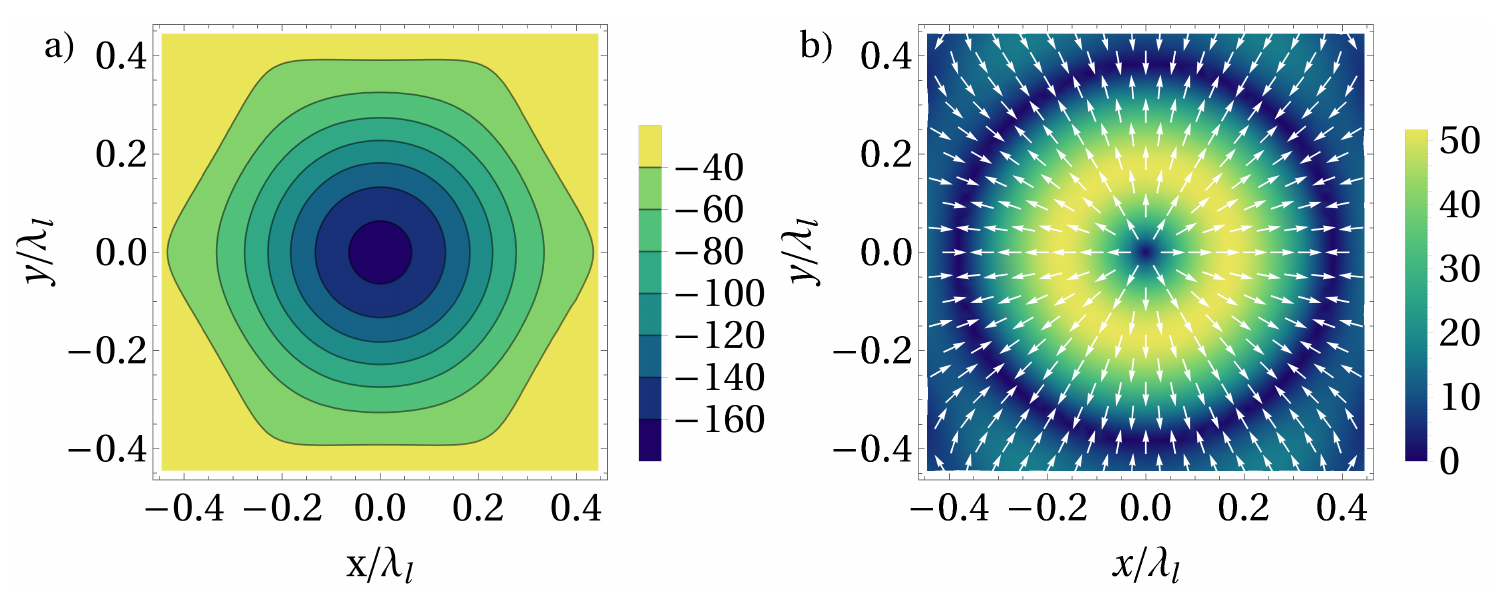}}
\caption{(a) Scalar potential $V(x,y)$ and (b) vector potential [fictitious magnetic field $\vec{B}_\mathrm{fic}(x,y)$], both in units of recoil energy ${\mathcal{E}}_{0} = \frac{(2\pi \hbar)^2}{2m \lambda_l^2}$.  The $z$ component of $\vec{B}_\mathrm{fic}$ vanishes, leaving only in-plane vector components. Both $V$ and $\vec{B}_\mathrm{fic}$ are radially symmetric to a very good approximation  close to the origin, $(0,0)$, but far from the center, both have hexagonal symmetry. Their magnitudes are proportional to the optical laser intensity.}
\label{fig:V_and_B_plot}
\end{figure}

We consider $^{87}$Rb atoms in a 2D SDOLP created by six laser beams, which are tightly focused in the $z$ direction, as proposed in Ref.~\cite{Kuzmenko_19}. The slowly varying electric field envelope is given by
$\vec{E}(x,y,t) = (E_0/3) \sum_{n=1}^6 \left( \vec{e}_z + \vec{q}_n \times \vec{e}_z \right) e^{i \vec{q}_n \cdot \vec{r} -i\omega_l t}$, where $\vec{e}_z$ is the unit polarization vector along the $z$-axis and the wave vectors are
$\vec{q}_n = - \frac{2\pi}{\lambda_l} (\cos(\frac{n \pi}{3}),\sin(\frac{n \pi}{3}),0)$ where $\lambda_l$ is the laser wavelength.  For $^{87}$Rb,  where the total electronic angular momentum is $J = 1/2$, the effective interaction of an atom with the electromagnetic field can be described using a scalar potential $V$ and fictitious magnetic field $\vec{B}_\mathrm{fic}$ \cite{Kuzmenko_19} (i.e., a vector potential), since the tensor terms vanish [see supplemental material (SM) \cite{SM}],
\begin{equation}  \label{H_Stark}
H_\mathrm{Stark} (x,y) = V(x,y) - g\mu_B \vec{B}_\mathrm{fic}(x,y) \cdot \vec{F},
\end{equation}
where $\vec{F}$ is the atomic hyperfine angular momentum, $\mu_B$ is the Bohr magneton, $g=1/2$, and
\begin{eqnarray}
V(x,y) &= -\frac{\alpha_0(\omega_l)}{4} \vec{E}^*(x,y) \cdot \vec{E}(x,y), \label{eq_for_V} \\
g\mu_B \vec{B}_\mathrm{fic}(x,y) &= i\frac{ \alpha_1(\omega_l)}{ 4(2I+1) } \vec{E}^*(x,y) \times \vec{E}(x,y).  \label{eq_for_B}
\end{eqnarray}
Here $\alpha_0$ and $\alpha_1$ are scalar and vector polarizabilities. Figure~\ref{fig:V_and_B_plot} shows full hexagonal potentials which are isotropic close to the site center. In the current study we focus on single site effects and assume that atomic wave function is well localized to only the center of the site, hence even though the lattice has hexagonal symmetry, potentials can be treated as isotropic (see Ref.~\cite{Kuzmenko_19} and the SM). Note that the divergence of the fictitious magnetic field does not vanish and $\vec{B}_\mathrm{fic}(r)$ corresponds to radially distributed magnetic monopole density.

First, we solve the Schr\"odinger equation for a single atom in a lattice cell in the presence of an additional uniform magnetic field $B_\mathrm{\rm{ext}}$ and find the energy eigenvalues and eigenfunctions.  The projections of ${\bf F}$, the orbital angular momentum of the atom about the SDOLP minimum ${\boldsymbol \ell}$, and the total angular momentum of the QR ${\boldsymbol {\mathcal{L}}}$ (where ${\boldsymbol {\mathcal{L}}} = {\bf F} + {\boldsymbol \ell}$), on the $z$-axis are $m_F$, $m_\ell$ and $\zeta = m_F + m_\ell$ respectively. Close to the potential minimum at $\vec{r}= (x,y) = (0,0)$ in a strong enough laser field the SDOLP is radially  symmetric, which allows to classify energy eigenstates using principal quantum number $n$ ($n=0,1,2 ...$) and projection of the total angular momentum on the $z$ axis $\zeta = 0,\pm1,\pm2,\dots$. For details of the single-atom system see Supplemental Material.

Next we model a BEC of $^{87}$Rb atoms in the $F=1$ state in the electromagnetic field generated by the lasers that form the SDOLP in the $x$-$y$ plane (the width of the SDOLP in the $z$-direction is the smallest length scale, which is of order of laser wavelength $\lambda_l$, our unit of length) in the presence of an external magnetic field transverse to the SDOLP. The BEC ground state wave function is also an eigenstate of ${\boldsymbol {\mathcal{L}}}_z$ with eigenvalue $\zeta$.  Both spin and orbital angular momentum contribute to $\zeta$, therefore it describes the topology of the BEC spinor components, i.e., their vortex structure. We show that an external magnetic field applied along the symmetry axis leads to a transition in which the ground state switches from $\zeta=0$ to $\zeta=1$ state as the external magnetic field varies. As both states are protected by the symmetry, this transition must be triggered by fluctuations that break the axial symmetry.

Atoms in the BEC interact by contact forces with both spin-dependent and spin-independent contributions, which are expressed by the coefficients $c_0$ and $c_2$ respectively, related to the scattering lengths $a_0$ and $a_2$, as follows: $c_0 = 4\pi \hbar^2 (a_0 + 2 a_2)/3m$ and $c_2 = 4\pi \hbar^2 (a_2 - a_0)/3m$. For $^{87}$Rb, $a_0 = 5.387$ nm and $a_2 = 5.313$ nm. The scattering length $a_0$ parameterizes the interactions where two colliding atoms have a total spin atomic angular momentum $F_\mathrm{tot} = 0$, while $a_2$ corresponds to $F_\mathrm{tot} = 2$ \cite{Ho,Stenger_98}. Since our calculations are 2D, we renormalize $c_0$ and $c_2$ by dividing them by the vertical extend of the BEC: $c_0,c_2 \rightarrow c_0/\lambda_l, c_2/\lambda_l$.  The Hamiltonian density for an $F=1$ spinor with linear Zeeman splitting induced by an external magnetic  field ${\bf B}_{\rm{ext}} = B_{\rm{ext}} \, \vec{e}_z$ is 
\cite{Ueda1, Ueda3, Stamper}
\begin{eqnarray}
\mathcal{H} &=
\frac{\hbar^2}{2 m} \mathbf{\nabla} \Phi_a^\dagger \cdot \mathbf{\nabla} 
\Phi_a+ 
\Phi_a^\dagger \left[ V(x,y) - g\mu_B \vec{B}_\mathrm{fic}(x,y) \cdot \vec{F} \right] \Phi_a  \nonumber \\
&+
\frac{c_0}{2} \Phi_a^\dagger \Phi_b^\dagger \Phi_b \Phi_a + \frac{c_2}{2} \Phi_a^\dagger \Phi_{a'}^\dagger \mathbf{F}_{ab} \cdot \mathbf{F}_{a'b'} \Phi_{b'} \Phi_b  \nonumber \\
&- g\mu_B B_{\rm{ext}} \Phi_a^\dagger (F_z)_{ab} \Phi_b .
\label{hamiltonian}
\end{eqnarray}
Here, $\Phi(\vec{r}) = (\Phi_1(\vec{r}),\Phi_0(\vec{r}),\Phi_{-1}(\vec{r}))^T$ is a vector composed of the field annihilation operators for an atom at point $\vec{r}$, in the hyperfine state components $m_F=1$, $0$, and $-1$ respectively, $m$ is the mass of the particles, ${\bf F}$ is the total spin atomic angular momentum vector and each component is a 3$\times$3 spin-1 matrix (with $F_z$ being a diagonal tensor), $B_\mathrm{\rm{ext}}$ is the magnitude of the (uniform) external magnetic field in the $z$-direction, and the last term is the linear Zeeman Hamiltonian \cite{Ueda1}.

\subsection{\textbf{\textit{BEC in a lattice site}}}

The $^{87}$Rb Bose-Einstein condensate in the hyperfine $F=1$ state is trapped in the SDOLP with tight confinement in the $z$ direction. The strength of the SDOLP is determined by the intensity of laser beams, which we take to be 70 W/cm$^2$, and the wavelength of the laser beams is taken as $\lambda_l=795$ nm. We assume that there is an additional transverse external magnetic field $B_\mathrm{\rm{ext}}$ in the $z$ direction whose strength can be tuned. Then, assuming the mean-field approximation applies, we use the imaginary-time technique \cite{ITM_Chiofalo} to find the ground state wave functions (order parameters) of a QR BEC.  The wave functions form a three component spinor $\Psi=(\psi_{+1},\psi_0,\psi_{-1})^T$. 

\begin{figure}
\centering
\includegraphics[width=\linewidth]{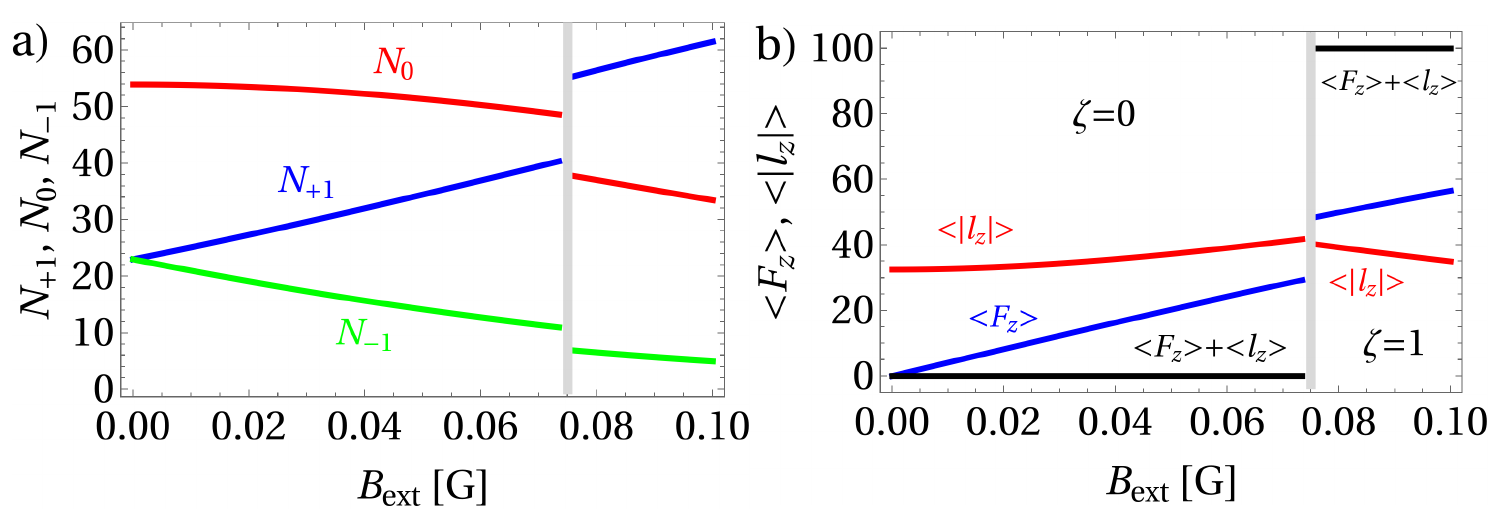}%
\caption{
(a) The number of atoms in each ground state QR BEC spinor component versus external magnetic field.  With increasing magnitude of the external magnetic field, atoms tend to accumulate in the $m_F=+1$ component at the expense of the $m_F=0$ and $m_F=-1$ components. Both $\zeta=0$  and $\zeta=1$ states behave similarly. (b) The mean value of $\langle F_z \rangle$, $\langle \ell_z \rangle$, and the sum $\langle F_z + \ell_z \rangle$ for the QR BEC ground state as a function of external magnetic field. $\langle F_z \rangle$ increases with external magnetic field due to atomic spin reorientation and alignment along the ${\bf e}_z$ axis. Simultaneously, $\langle |\ell_z| \rangle$ increases with external magnetic field  in the $\zeta=0$ state while significantly decreasing in the $\zeta=1$ state, compare with Fig.~\ref{Density of BEC at 6080mG}. The black horizontal line shows a mean value of the $z$-component of the total angular momentum, $\hbar \zeta = \langle F_z + \ell_z \rangle$ and conservation of $\zeta$ results from axial symmetry of the Hamiltonian.}

\label{fig:number of atoms}
\end{figure}

We set the total number of atoms to $N=100$, and vary the external magnetic field strength in the range from $0$ mG to $100$ mG. In Fig.~\ref{fig:number of atoms} we follow the ground state of the QR as the external magnetic field is increased, and plot the number of atoms in each spinor component as a function of external magnetic field. With $B_\mathrm{\rm{ext}} = 0$, the $m_F=0$ component is the most populated, while equal number of atoms occupy the $m_F=\pm 1$. This distribution is selected to minimize interaction with fictitious magnetic field. As the external magnetic field increases, atoms move from the $m_F=-1$ and $m_F=0$ components to the $m_F=+1$. This happens since the energy of a magnetic dipole gets lower when a dipole moment is oriented in the same direction as an external magnetic field. In our case $B_\mathrm{\rm{ext}}$ is oriented in the $+\vec{e}_z$ direction, which favors the $m_F=+1$ component. As we can see from Fig.~\ref{fig:number of atoms}, for a certain value of $B_\mathrm{\rm{ext}}$ the number of atoms and atomic angular momentum as a function of $B_\mathrm{\rm{ext}}$ is not continuous, which is an indication of an abrupt transition. After this transition the occupation of the $m_F=+1$ component is the largest, and there are very few atoms left in the $m_F=-1$. In the lower panel of Fig.~\ref{fig:number of atoms} we present $z$ component of the total angular momentum, atomic hyperfine spin $\langle F_z \rangle$, and the modulus of the atomic orbital angular momentum, $\langle |\ell_z| \rangle$, calculated for the full three-component spinor as a function of external magnetic field. Orbital angular momentum per atom in $m_F=+1,0,-1$ spinor components before the transition is equal to $m_\ell=-1,0,1$ and after the transition it changes to $m_\ell=0,1,2$, respectively.

\begin{figure}[htb]
\centering
\includegraphics[width=1\linewidth]{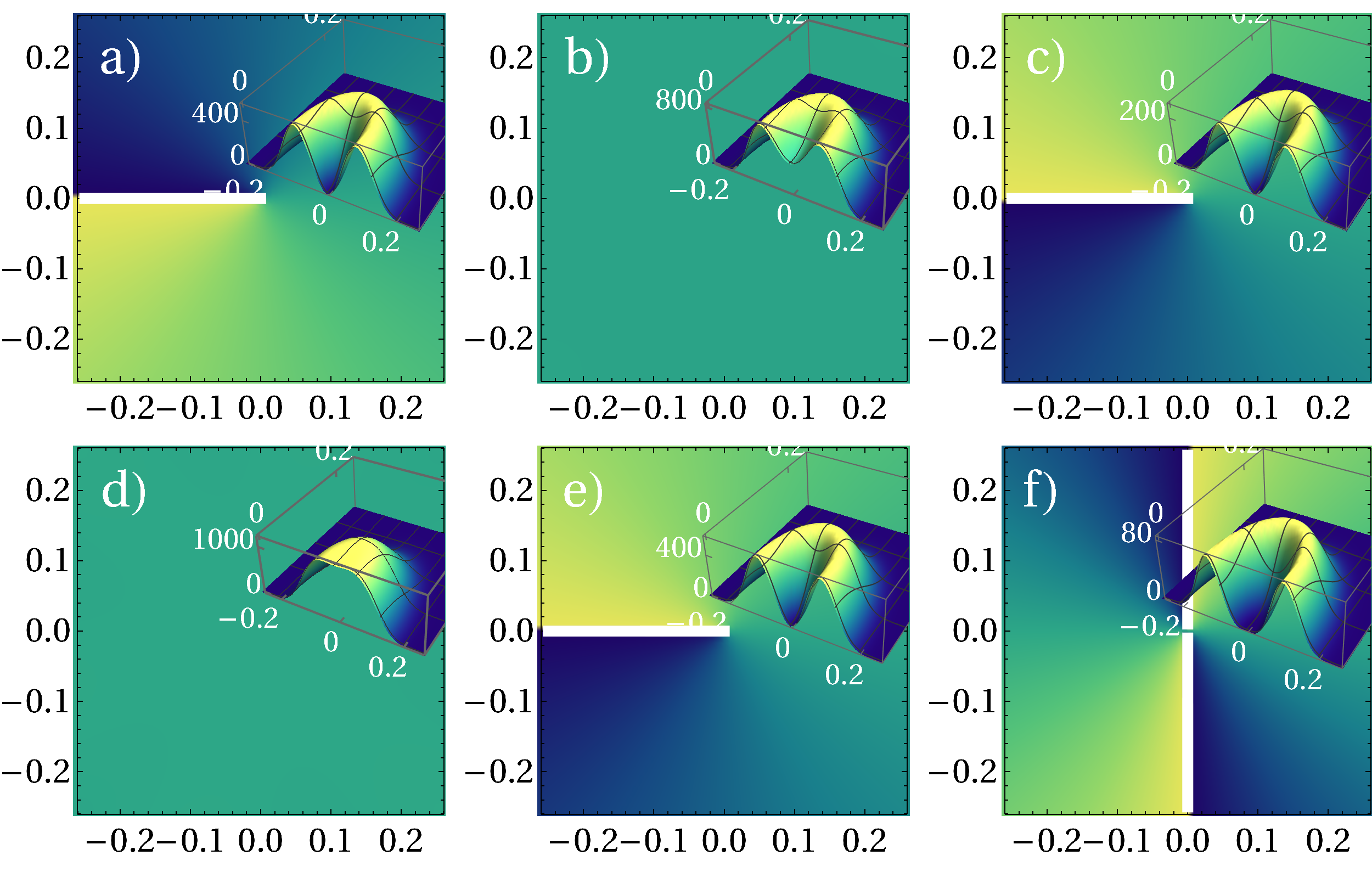}
\caption{Phase and density (see insets) of each spinor component of the ground state QR BEC at $B_\mathrm{\rm{ext}}=40$ mG ($\zeta=0$, top row) and the ground state at $B_\mathrm{\rm{ext}}=100$ mG ($\zeta=1$, bottom row). The units of the axes are the same units as in Fig.~\ref{fig:V_and_B_plot}. To show behavior of the densities at the origin we plot only half of the profiles by cutting them along $y=0$. The left panel is for  $m_F=+1$, middle $m_F=0$ and right $m_F=-1$. Phases and densities indicate that there are vortices with charge -1, 0 and $1$ in $\zeta=0$ (upper row) and 0, 1 and 2 in the $\zeta=1$ (lower row).} 
\label{Density of BEC at 6080mG}
\end{figure}

\begin{figure}
\centering
\includegraphics[width=1\linewidth]{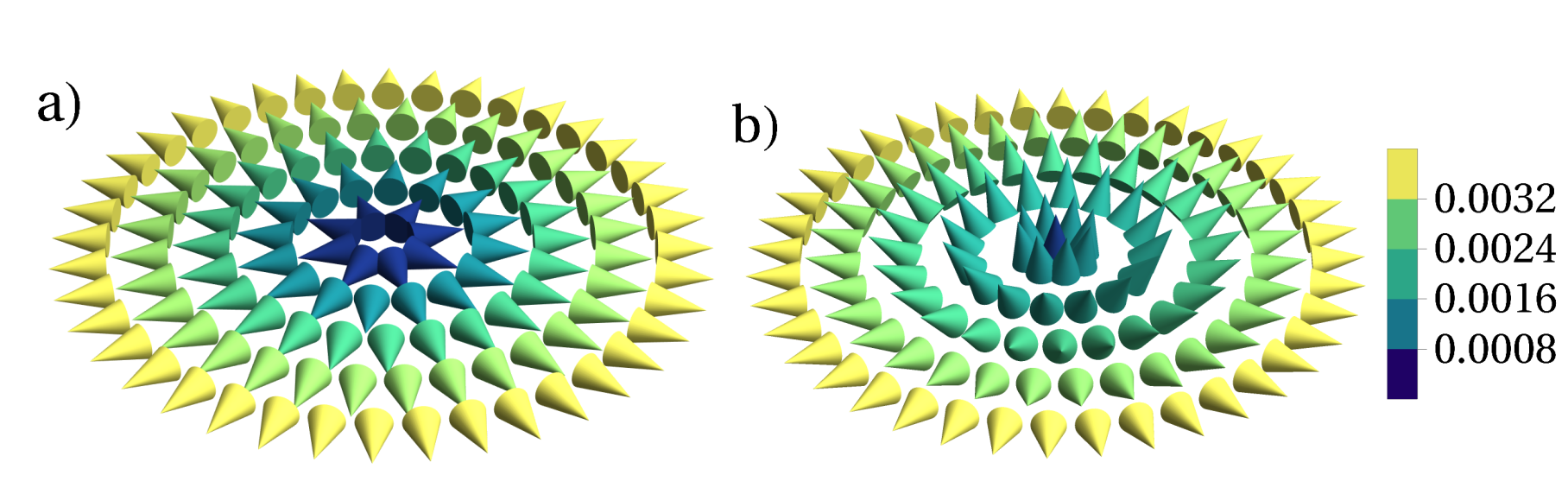}
\caption{$\langle \vec{F}(x,y) \rangle$ for two different external magnetic field magnitudes in the region $x,y \in [0,0.077\,\lambda_l]$. (a) The state $\zeta = 0$ for $B_\mathrm{\rm{ext}}=40$ mG and (b) the state $\zeta = 1$ for $B_\mathrm{\rm{ext}}=100$ mG. The color coding is used to specify the length of the spin (in units of $\hbar$ per atom) and the direction depicts the orientation of the spin. With increasing external magnetic field, the atomic spin increasingly aligns along the ${\bf e}_z$ axis.}
\label{fig:3D plot spin}
\end{figure}

In the following we focus on differences in the characteristics of atom distributions (Fig.~\ref{Density of BEC at 6080mG}) and spin distributions (Fig.~\ref{fig:3D plot spin}) before and after transition. Ground state density and the phase for all three spinor components  are shown in Fig.~\ref{Density of BEC at 6080mG}, where the left column corresponds to the value of external magnetic field equal to $B_\mathrm{\rm{ext}}=40$ mG. We see that the density of the $m_F=+1$ component at the position $\vec{r}=(0,0)$ for this value of the magnetic field is equal to zero. The same is true for $m_F=-1$, but not in $m_F=0$, where only a shallow minimum at $\vec{r}=(0,0)$ is present.  All density and phase profiles are in compliance with the expectation values of the orbital angular momentum per atom about the minimum of the SDOLP (${\boldsymbol \ell}=\vec{r}\times\vec{p}$), which is equal to $-\hbar,0,+\hbar$, for the $m_F=+1,0,-1$ spinor components of the ground state wave function at this value of $B_\mathrm{\rm{ext}}$. 

In the right column of  Fig.~\ref{Density of BEC at 6080mG} we present analogous results for the value of external magnetic field $B_\mathrm{\rm{ext}}=100$ mG, above the transformation. The difference is evident since now the $m_F=+1$ state has a Gaussian-like shape, which together with the phase structure indicates no vortex in this component. We calculated the expectation value of the atomic orbital angular momentum for each component and this time obtained values $0, \hbar, 2\hbar$ for the $m_F=+1,0,-1$ respectively. Comparing states (densities and phases) before and after the transition we see increase of angular momentum per atom by $\hbar$ in each component. It is only possible if the components in left column belong to the $\zeta=0$ and in the right column to $\zeta=1$ states. Regardless of the distribution of atoms among the three spinor components, the expectation value of $\zeta=m_F+m_\ell$ per atom is equal to zero in the left column and one in the right column of Fig.~\ref{Density of BEC at 6080mG}. So at each particular value of $B_\mathrm{\rm{ext}}$ it is the same in each $m_F$ component, because of symmetry.  This is in fact the analogue of the celebrated Einstein-de Haas effect \cite{EdeHaas}, a manifestation of the fact that spin contributes to the total angular momentum on the equal footing with the rotational angular momentum. This explains why the radical change of the ground state topology occurs (see the discussion below).

In Fig.~\ref{fig:3D plot spin} we portray the spin textures by constructing vectors of expectation values $\langle F_x \rangle, \langle F_y \rangle$, and $\langle F_z \rangle$ in a ground state spinor wave function.  The results are shown as a 3D vector plot $\langle \vec{F}(x,y) \rangle$. Before the transition, the spin vector $\vec{F}$ mainly lies in the $x$-$y$ plane. After the transition the spin texture changes. In Fig.~\ref{fig:3D plot spin}(b), the spin vector $\vec{F}$ in the vicinity of $\vec{r}=(0,0)$, where the atoms are mostly located, is turned towards the $z$ direction, trying to along with the external magnetic field. This happens  because in this region the external magnetic field is stronger than the vector optical potential. After the transition, and further from the center, the spin vector $\vec{F}$ is still aligned with the vector part of the optical potential. Reorientation of $\vec{F}$ is a clear signature of the change of the ground state of the QR BEC.

\section{Origin of the ground state topological transition}
\label{sec:origin}

\begin{figure}[bth]
\centering
\includegraphics[width=0.95\linewidth]{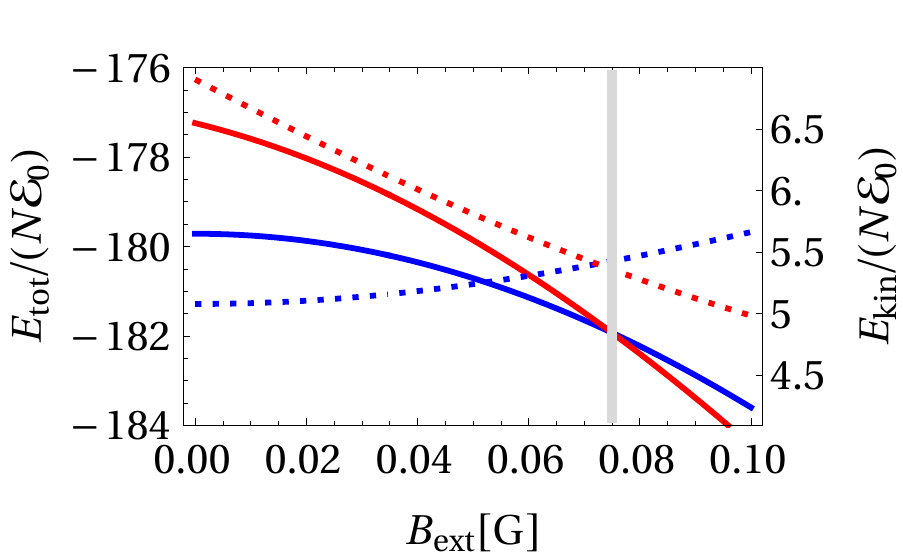}
\caption{Total (solid) and kinetic (dashed) energies for the two lowest QR states, $\zeta=0$ (blue lines) and $\zeta=1$ (red lines) versus external magnetic field strength. The gray vertical line emphasizes that for two lowest QR states both total and kinetic energies cross at approximately the same external magnetic field magnitude.}
\label{crossingenergies}
\end{figure}

To understand the origin of the transition we take a closer look at the behavior of energy contributions for two lowest QR states. In Fig.~\ref{crossingenergies} we plot the total (solid lines) and kinetic (dashed lines) energies as a function of an external magnetic field for states with indices $\zeta=0$ (blue) and $\zeta=1$ (red). The crossing of two solid lines (total energy curves) occurs approximately at $B_{\mathrm{ext}}=70\,$mG at the intensity of $I=70$W/cm$^2$.
For a hexagonal lattice geometry with a large number of atoms ($N=100$) per lattice site we checked that the transformation occurs at the critical value of $B_{\mathrm{ext}}/B_{\mathrm{fic}}=0.7$, where $B_{\mathrm{fic}}$ is the magnitude of the fictitious magnetic field, Eq.~(\ref{eq_for_B}), by varying the strength of the SDOLP potential (and therefore the strength of the fictitious magnetic field) over a wide range. We know from Fig.~\ref{Density of BEC at 6080mG} that the topology of the ground state is changed after the crossing point, and the winding number of each spinor component is increased by one.

When the ratio of $B_{ext}/B_{fic}$ field is low atomic spin are mostly in $x-y$ plain and predominantly occupy $m_F=0$ component, minimizing both kinetic and magnetic energies. Above the critical ratio, atoms in the ground state are moved to the $m_F=+1$ component to minimize magnetic energy. However, the topology of the $m_F=+1$ component is different for $\zeta=0$ and $\zeta=1$ states; the former carries `charge' $-1$ vortex, while the latter has no vortex at all. There is an extra rotational kinetic energy in a vortex state, hence transferring atoms to the $m_F=+1$ component results in the increase of the kinetic energy in $\zeta=0$ state and decrease in the $\zeta=1$ state. These two kinetic energies cross in Fig.~\ref{crossingenergies} and the crossing point is located in the close vicinity to the crossing of total energies. Since all the other energy components (Zeeman and SDOLP field) of two lowest QR states change with external magnetic field in a similar way for both $\zeta=0$ and $\zeta=1$ states, it is exactly kinetic energy that is responsible for the transition. For this transition to occur a symmetry breaking perturbation is required.

\section{Conclusions}

We proposed a realization of a quantum rotor system, based on a BEC of $^{87}$Rb atoms in a 2D spin-dependent optical lattice potential. When enriched by adding a transverse external magnetic field, the system exhibits a topological transition of the ground state as a function of the strength of the external magnetic field. One observes the change of the topology of the quantum rotor BEC ground state; it switches from the one at low magnetic fields corresponding to the rotor angular momentum $\zeta=0$, to $\zeta=1$ at higher field strengths. The reason for this change is related to the transfer of atoms triggered by the external magnetic field, which combined with the Einstein-de Haas effect, leads to the subtle interplay between various energy components in the two lowest quantum rotor states. These two distinct ground states can be observed in experiment by measuring the density profiles of different components following Stern-Gerlach separation.  Extensions of this work to consider both sides of the superconductor-insulating transition regimes and other spin-dependent optical lattice potential geometries are in progress. Although in this study we do not consider dynamics, but rather the states of the system at specific magnetic field strengths, it may be useful to study dynamics as the magnetic field strength varies in time.

\section{Acknowledgments}

MT was supported by the Polish National Science Center (NCN) Grant 2016/22/M/ST2/00261. MB and MG were supported by the NCN Grant No. 2019/32/Z/ST2/00016 through the project MAQS under QuantERA, which has received funding from the European Union’s Horizon 2020 research and innovation program under grant agreement no 731473.
PS was supported in part by the National Science Centre (NCN), grant no. 2017/25/B/ST2/01943.

\newpage

{\bf Topological phase transition for a quantum rotor Bose-Einstein condensate: Supplemental material}

In this Supplemental Material for the main text (MT) \cite{main_SM} we discuss the effective interaction between a single atom and the light generating a spin-dependent optical lattice potential with hexagonal symmetry.

\subsection{$^{87}$Rb fine and hyperfine structure}

The energy levels of Rb are split due to the interaction between the orbital angular momentum $\vec{L}$ with the spin angular momentum for the electrons, and the resulting energy level splitting is called fine structure. The total electronic angular momentum,
\begin{equation}
\vec{J}=\vec{L}+\vec{S} ,
\end{equation}
has corresponding quantum number $J$ that satisfies the inequality
$|L-S|\le J \le |L+S|$.
Operator $\vec{J}^2$ has an eigenvalue $J(J+1)\hbar^2$ and the operator $J_z$ has $2J+1$ eigenvalues $\hbar m_J$ with $m_J \in \{ -J, -J+1, \dots, J \}$. The same applies for operators $\vec{L}$ and $\vec{S}$. For convenience, we shall sometimes set $\hbar$ equal to unity.
Rb also has hyperfine energy level splitting due of the coupling between the total electronic angular momentum $\vec{J}$ with the nuclear spin operator $\vec{I}$. We denote the quantum number corresponding to the total atomic angular momentum operator $\vec{F}$ as $F$, where
\begin{equation}
\vec{F} = \vec{J} + \vec{I}.
\end{equation}
The total atomic angular momentum quantum number $F$ satisfies inequality
$|J-I|\le F \le |J+I|$.
The nuclear spin of $^{87}$Rb is $I = 3/2$, which combined with $J = 1/2$ yields $F = 1$ and $F = 2$.  The ground hyperfine state has $F = 1$.

\subsection{AC Stark shift}
For an atom with total electronic angular momentum $J$, the light-atom interaction can be calculated by second-order perturbation theory.  The resulting ac Stark potential can be written as \cite{LeKien_13_SM, Kuzmenko_19_SM}
\begin{align}
U(\vec{r})  &= -\frac{1}{4} \left\{ \alpha_{n,J}^s(\omega_l) \vec{E}^*(\vec{r})\cdot\vec{E}(\vec{r}) \right. \nonumber \\
&- \left. \frac{i\alpha_{n,J}^v(\omega_l)}{2J}\left[\vec{E}^*(\vec{r})\times\vec{E}(\vec{r})\right]\cdot\vec{J}     \right.\\
&+ \frac{3\alpha_{n,J}^t(\omega_l)}{2J(2J-1)} \left[ (\vec{E}^*(\vec{r})\cdot\vec{J})(\vec{E}(\vec{r})\cdot\vec{J}) \right. \nonumber \\
&+\left.(\vec{E}(\vec{r})\cdot\vec{J})(\vec{E}^*(\vec{r})\cdot\vec{J}) \right.\\
&\left.\left.-\frac{2}{3}J(J+1)(\vec{E}^*(\vec{r})\cdot\vec{E}(\vec{r})\right]\right\},
\end{align}
where $\vec{E}(\vec{r})$ is the spatial part of the complex electric field envelope, and $\alpha_{n,J}^s(\omega_l)$, $\alpha_{n,J}^v(\omega_l)$, $\alpha_{n,J}^t(\omega_l)$ are the scalar, vector and tensor polarizabilities of an atom in fine-structure level $\left|nJ\right>$, $n$ being the principal quantum number. From Ref.~\cite{LeKien_13_SM},
\begin{align}
\alpha_{n,J}^s(\omega_l) &= \frac{\alpha_{n,J}^{(0)}(\omega_l)}{\sqrt{3(2J+1)}},\\
\alpha_{n,J}^v(\omega_l) &= -\frac{\sqrt{2J}\alpha_{n,J}^{(1)}(\omega_l)}{\sqrt{(J+1)(2J+1)}},\\
\alpha_{n,J}^t(\omega_l) &= -\frac{\sqrt{2J(2J-1)}\alpha_{n,J}^{(2)}(\omega_l)}{\sqrt{3(J+1)(2J+1)(2J+3)}},
\end{align}
where $\alpha_{n,J}^{(K)}(\omega_l)$, for $K = \{0,1,2\}$ is
\begin{align}
\alpha_{n,J}^{(K)}(\omega_l) &= (-1)^{K+J+1} \sqrt{2K+1}\nonumber\\
&\sum_{n',J'} (-1)^{J'}
\begin{pmatrix}
1 & K & 1 \\
J & J' & J \\
\end{pmatrix}
\left| \left< n'J'||\vec{d}||nJ \right> \right|^2   \nonumber
\end{align}
\begin{align}
\times\, \frac{1}{\hbar}\mathrm{Re}&\left(
\frac{1}{\omega_{n',J'\to n,J} -\omega_l -i\gamma_{n',J'\to n,J}/2 }\right. \nonumber\\
&+\left.\frac{(-1)^K}{\omega_{n',J'\to n,J} +\omega_l +i\gamma_{n',J'\to n,J}/2 }
 \right).
\end{align}
Here $\omega_{n',J'\to n,J}$ is a excitation frequency from $\left|nJ \right>$ to $\left|n'J' \right>$ fine-structure level,
$\begin{pmatrix}
1 & K & 1 \\
J & J' & J \\
\end{pmatrix}$
is the Wigner 6-$j$ symbol and $\left< n'J'||\vec{d}||nJ \right>$ is the reduced matrix element of the dipole moment operator.

\subsection{Spin-dependent optical lattice potential}

We explicitly consider a spin dependent optical lattice potential (SDOLP) with hexagonal symmetry, as proposed in Ref.~\cite{Kuzmenko_19_SM}.  Six laser beams, which are tightly focused in the $z$ direction, are the source of the electromagnetic field. The slowly varying electric field envelope $\vec{E}(x,y)$ is given by
$\vec{E}(x,y,t) = \frac{1}{3} \sum_{n=1}^6 \left( \vec{e}_z + \vec{q}_n \times \vec{e}_z \right) e^{2\pi i \vec{q}_n \cdot \vec{r} -i\omega_l t}$
where $\vec{e}_z$ is the unit polarization vector along the $z$-axis and the wave vectors are
$\vec{q}_n = -(\cos(\frac{n \pi}{3}),\sin(\frac{n \pi}{3}),0)$.  In the dipole approximation, the light-atom interaction $\mathcal{V}_{la}$ is given by the operator
\begin{equation}
\mathcal{V}_{la} = -\vec{E}\cdot\vec{d} = -\frac{1}{2} \left( E_0 \vec{u} e^{-i\omega_l t} + \mathrm{c.c.} \right) \cdot \vec{d} ,
\end{equation}
where $\vec{d}$ is the transition dipole operator.  The second-order ac Stark shift (second order in $\mathcal{V}_{la}$) for a non-degenerate atomic energy level $\left| a \right>$ is
\begin{align}
\delta E_a = -\frac{|E_0|^2}{4\hbar} \sum_b \mathrm{Re} \left( \frac{|\left<b|\vec{u}\cdot\vec{d}|a\right>|^2}{\omega_b-\omega_a-\omega_l-i\gamma_{ba}/2} + \right. \nonumber \\ 
\left. + \frac{|\left<a|\vec{u}\cdot\vec{d}|b\right>|^2}{\omega_b-\omega_a+\omega_l+i\gamma_{ba}/2} \right).
\end{align}
The ac Stark interaction operator for an atom with electronic angular momentum $J$ can then be written using multipole notation as
\begin{align} \label{Eq:W}
W=-\frac{1}{4} |E_0|^2 \left\{ \alpha_{nJ}^s(\omega_l) - i\alpha_{nJ}^v(\omega_l) \frac{[\vec{u}^*\times\vec{u}]\cdot\vec{J}}{2J} +  \right. \nonumber \\
\left. + \alpha_{nJ}^T(\omega_l) \frac{3[(\vec{u}^*\cdot\vec{J}) (\vec{u}\cdot\vec{J})
(\vec{u}\cdot\vec{J}) (\vec{u}^*\cdot\vec{J}) ] -2\vec{J}^2 }{2J(2J-1)} \right\}.
\end{align}
The total internal atomic angular momentum $\vec{F}$ is the vector sum of the total electronic angular momentum $\vec{J}$ and the nuclear spin $\vec{I}$. In case of $^{87}$Rb, $I$ is equal to $3/2$ and $J$ is equal to $1/2$. Therefore, we have two atomic hyperfine states with $F$ equal to $1$ or to $2$, and the ground hyperfine states has $F=1$. Reference \cite{Kuzmenko_19_SM} considered ${}^6$Li, which is a fermion, in a SDOLP with one atom per lattice site.

For $J = 1/2$, the effective interaction of an atom with the electromagnetic field can be described using a scalar potential $V$ and a vector potential which is often called a fictitious magnetic field $\vec{B}_\mathrm{fic}$ \cite{Kuzmenko_19_SM}.  The tensor term vanishes, hence the ac Stark Hamiltonian, obtained via second order perturbation theory in the light-atom interaction takes the form given in Eq.~(2) of the MT, $H_\mathrm{Stark} (x,y) = V(x,y) - g\mu_B \vec{B}_\mathrm{fic}(x,y) \cdot \vec{F}$, where $g=1/2$. The scalar potential $V$ and the fictitious magnetic field $\vec{B}_\mathrm{fic}$, which appears in the second term on the right-hand side, are given by Eqs.~(3) and (4) in the main text, where the $\alpha_0$ and $\alpha_1$ are scalar and vector terms of the atomic polarizability and are given in the hyperfine basis (the details of the derivation can be found in Ref.~\cite{LeKien_13_SM}). It is worth noting that for atoms with $J$ greater than $1/2$, the polarizability also contains a non-vanishing tensor term.  Both $V$ and $\vec{B}_\mathrm{fic}$ are proportional to the laser intensity. The higher the laser intensity, the stronger an atom is attracted to the high intensity area of the potential when the SDOL lasers are red-detuned from resonance, i.e., atoms are trapped near the center of a lattice cell.  While the depth of the potential depends on the laser intensity, the size of the cell of the optical lattice potential does not. The size of the SDOLP cell depends on the laser wavelength.
We plot the scalar potential $V$ and the fictitious magnetic field $\vec{B}_\mathrm{fic}$ for $^{87}$Rb as a function of space within one lattice site in Fig. 1 in the MT.  The potential and the fictitious magnetic field are periodic and form a lattice consisting of identical hexagonal cells.

If an external magnetic field ${\bf B}_{\rm{\rm{ext}}}$ is present, a Zeeman Hamiltonian to account for the interaction of the atom with the external field must be added to the Hamiltonian due to the SDOLP.  Below we consider only cases where the external magnetic field is along the $z$-axis, i.e., ${\bf B}_{\rm{\rm{ext}}} = B \, \vec{e}_z$.
  
\subsection{Scalar and vector terms of the atomic polarizability}

The first excited state of $^{87}$Rb is a fine structure doublet, i.e., it splits into two states $5\,{}^2P_{1/2}$ and $5\,{}^2P_{3/2}$. The excitation wavelengths from the ground state, $5\,{}^2S_{1/2}$, are \cite{Rb87_line_data_SM}
\begin{align}
5 \, {}^2S_{1/2} \to 5 \, {}^2P_{1/2}, \qquad \lambda_{1/2} = 794.979 \, \mathrm{nm},\\
5 \, {}^2S_{1/2} \to 5 \, {}^2P_{3/2}, \qquad \lambda_{3/2} = 780.241 \, \mathrm{nm}.
\end{align}
Reduced matrix elements $d_J=|\left< n, J|| \vec{d} || n, 1/2 \right>|$ of the atomic dipole moment operator $\vec{d}$ for the above transitions are
\begin{equation}
d_{1/2} = 2.992 \, e a_B, \qquad d_{3/2} = 4.227 \, e a_B,
\end{equation}
where $e$ is the elementary charge and $a_B$ is the Bohr radius.  The scalar $\alpha_0(\omega_l)$ and the vector $\alpha_1(\omega_l)$ atomic polarizability terms of $^{87}$Rb are given by \cite{LeKien_13_SM},
\begin{align}
\alpha_0(\omega_l) &= \frac{d_{1/2}^2}{6 \hbar} \frac{1}{ \omega_{1/2} -\omega_l } + \frac{d_{3/2}^2}{6 \hbar} \frac{1}{ \omega_{3/2} -\omega_l }, \\
\alpha_1(\omega_l) &= \frac{d_{1/2}^2}{3 \hbar} \frac{1}{ \omega_{1/2} -\omega_l } - \frac{d_{3/2}^2}{6 \hbar} \frac{1}{ \omega_{3/2} -\omega_l }.
\end{align}
The atomic polarizability is a function of the laser frequency $\omega_l$. Figure~\ref{fig:polarizability} plots the scalar and vector terms of the polarizability for a laser red detuned from the $D_1$ line, where both scalar and vector terms are positive. This leads to an interaction that pulls an atom into a high intensity area of the scalar potential. Positive vector term of the polarizability leads to a fictitious magnetic field that is directed radially outward close to the center, but for larger distances from the center of a cell, the radial symmetry is broken.

\begin{figure}[h]
\centering
\includegraphics[width=0.95\linewidth]{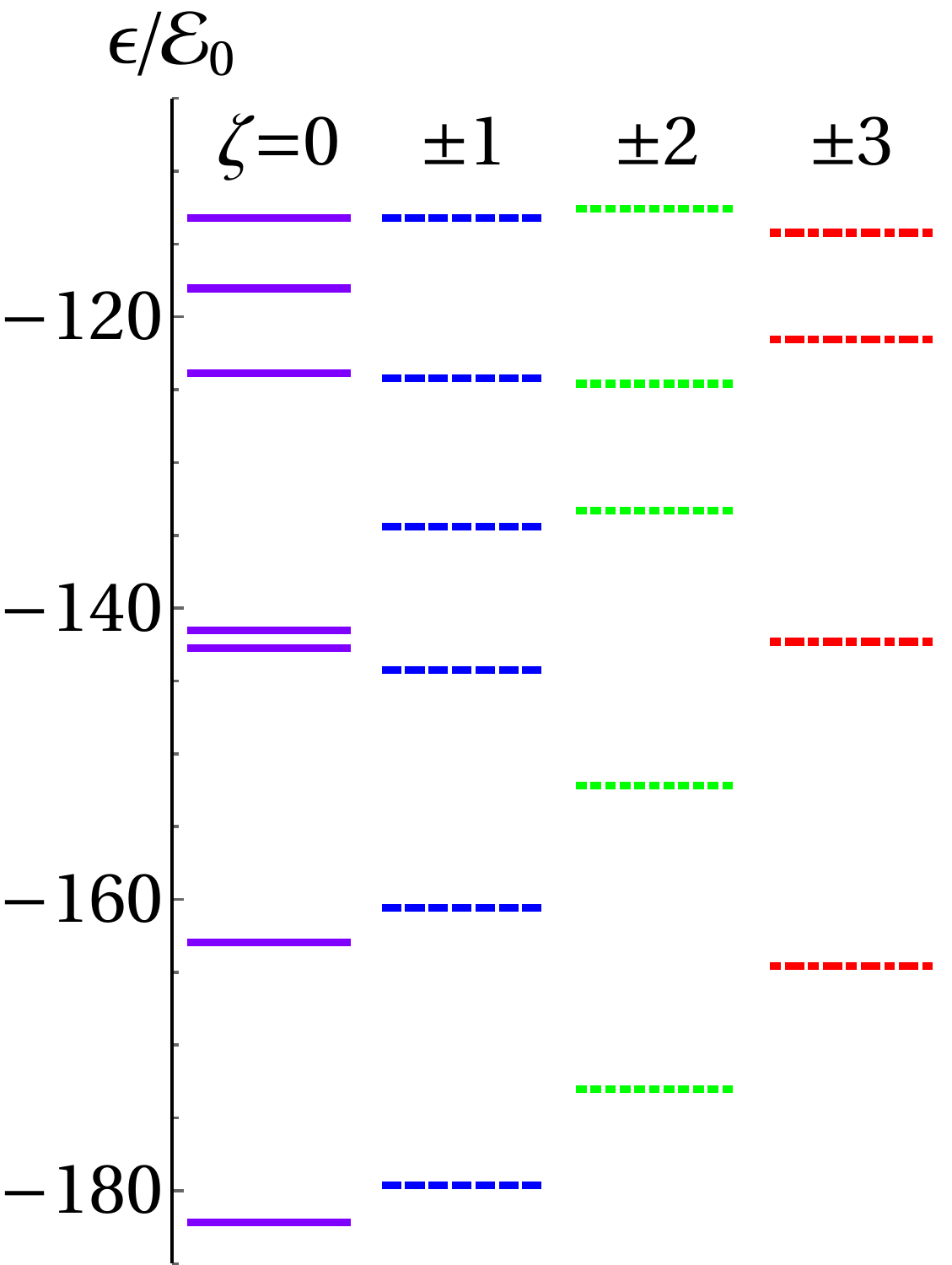}
\caption{Energy eigenvalues for a single $^{87}$Rb atom in SDOLP lattice cell in the isotropic approximation for the SDOLP.}
\label{fig:Single_atom_energy_levels zeta}
\end{figure}

\begin{figure}
\centering
\subfigure[]
{\includegraphics[width=0.9\linewidth]{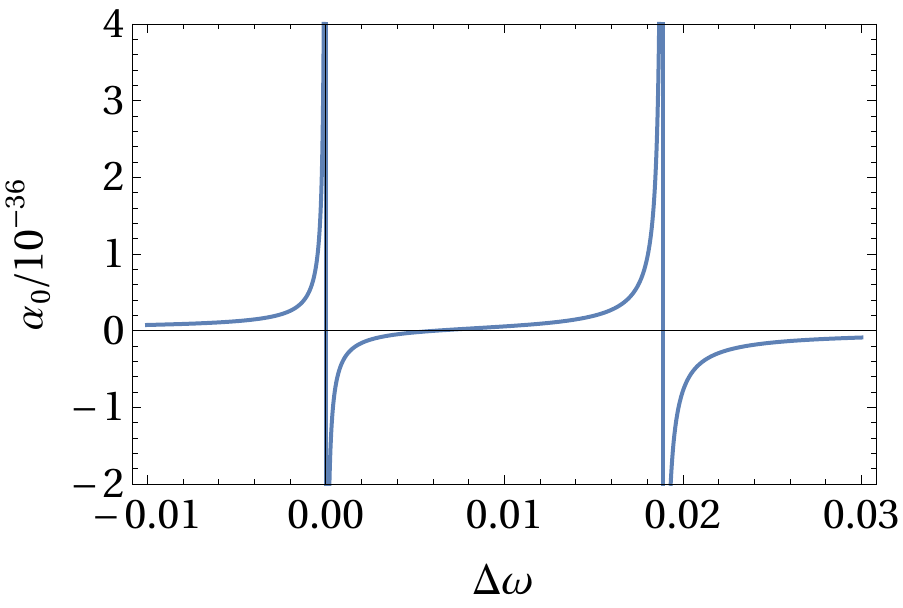}}
\qquad
\subfigure[]
{\includegraphics[width=0.9\linewidth]{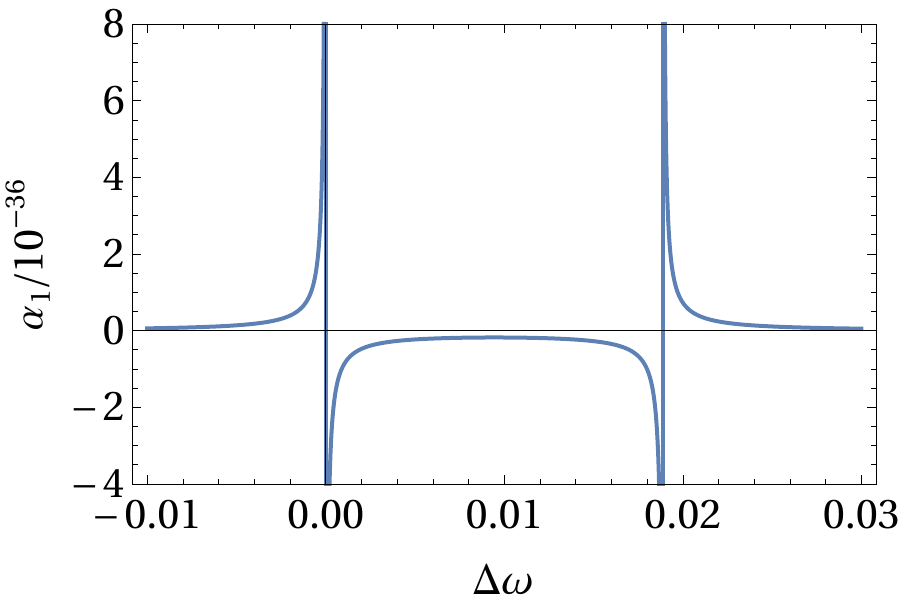}}
\caption{(a) Scalar polarizibility $\alpha_0(\omega_l)$, and (b) the vector polarizability $\alpha_1(\omega_l)$ as a function of laser frequency $\omega$ calculated for $^{87}$Rb. Both $\alpha_0$ and $\alpha_1$ have two resonant frequencies corresponding to the two fine structure excitation energies. The $D_1$ line is on the left, and the $D_2$ line is on the right. For laser red detuned from the $D_1$ line, the scalar term is positive, therefore a $^{87}$Rb atom is attracted to the high intensity area of the scalar potential. The vector polarizability is also positive for the laser red detuned from the $D_1$ line, therefore the fictitious magnetic field vector is oriented radially outward close to the center.}
\label{fig:polarizability}
\end{figure}

\begin{figure}
\centering
{\includegraphics[width=0.9\linewidth]{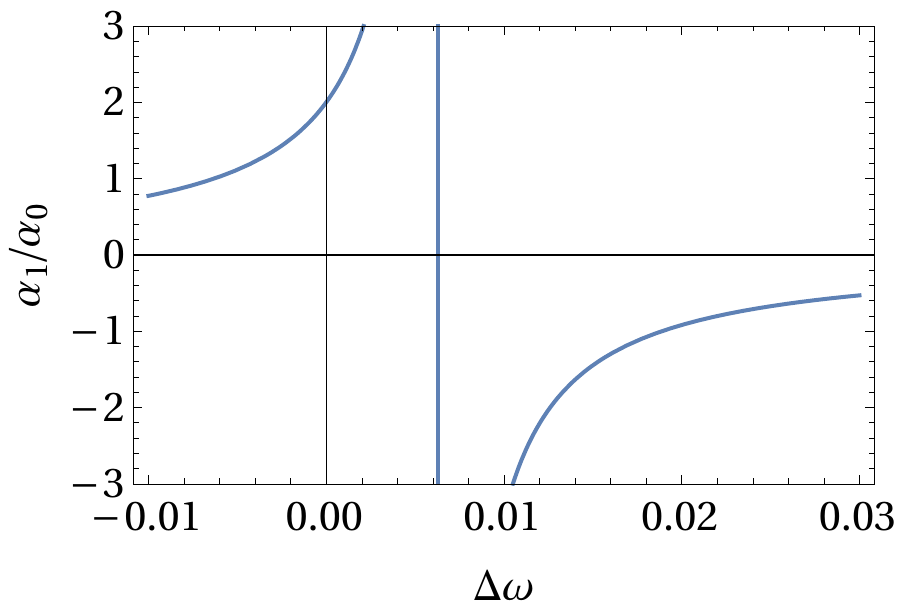}
}
\caption{Ratio of the vector and scalar polarizabilities, $\alpha_1(\omega_l)/\alpha_0(\omega_l)$. At the $D_1$ frequency, the ratio equals $2$.}
\label{fig:ratio_plot}
\end{figure}

We consider a laser that is red detuned from the $D_1$ line, at a wavelength of $\lambda_l = 795.456$ nm. The laser frequency was taken to be $\omega_l = 2.36801 \times 10^{15}$ rad/s. The ratio of the vector to the scalar polarizability at the laser frequency is $\frac{\alpha_1(\omega_l)}{\alpha_0(\omega_l)} = 1.82637$ (see Fig.~\ref{fig:ratio_plot}).


\subsection{Single atom in a lattice site: Isotropic approximation}

Close to the center of a SDOLP lattice site, say at  $\vec{r}=(0,0)$, the scalar and vector part of the SDOLP are isotropic. Following Ref.~\cite{Kuzmenko_19_SM}, we write approximate (i.e., radially symmetric) equations for the scalar potential $V(r)$
\begin{equation}
V = -\frac{\alpha_0(\omega_l) E_0^2}{6} \left[ 2+3J_0\left(\frac{2\pi r}{\lambda_l}\right) + J_0\left(\frac{2\sqrt{3}\pi r}{\lambda_l}\right)  \right],
\end{equation}
and for the fictitious magnetic field $\vec{B}(r)$, which has only a radial component
\begin{equation}
\vec{B} = \hat{{\bf r}} \, \frac{\alpha_1(\omega_l) E_0^2}{3(2I+1)} 
[
J_1( \frac{2\pi r}{\lambda_l}) +J_1( \frac{4\pi r}{\lambda_l}) +\sqrt{3}J_1( \frac{2\sqrt{3}\pi r}{\lambda_l})],
\end{equation}
where $E_0^2=2I/(c\varepsilon_0)$ and $\varepsilon_0$ is electric constant. In the isotropic approximation, $\zeta$ is a good quantum number (see Table \ref{quantum numbers}). We find solutions for a given $\zeta$. To do so, we write the Schr\"odinger equation for a single atom \cite{Kuzmenko_19_SM} in a spinor basis of eigenvectors of the spin operator component $F_r$,
\begin{align}
\chi_{+1}(r,\phi)&=\frac{1}{2} \left( e^{-i\phi}, \sqrt{2}, e^{i\phi} \right),\\
\chi_{0}(r,\phi)&=\frac{1}{\sqrt{2}} \left( e^{-i\phi}, 0, -e^{i\phi} \right),\\
\chi_{-1}(r,\phi)&=\frac{1}{2} \left( e^{-i\phi}, -\sqrt{2}, e^{i\phi} \right),
\end{align}
where the eigenvectors $\chi_{+1}(r,\phi),\chi_{0}(r,\phi),\chi_{-1}(r,\phi)$ are written in a basis of $\vec{F}_z$ eigenvectors. We can write the eigenvalue problem, $\mathcal{H} \Psi = E \Psi$, in a basis $\{\chi_i(r,\phi)\}_i$, by multiplying by an inverse of a matrix $U_\chi=(\chi_{+1}(r,\phi),\chi_{0}(r,\phi),\chi_{-1}(r,\phi))^\mathrm{T}$ from the left, and replacing the three-component spinor wave function $\Psi$ with $\Psi_\chi$, so
\begin{equation}
\Psi_\chi = \frac{1}{\sqrt{2\pi r}} e^{i\zeta\phi} U_\chi^{-1} \Psi,
\end{equation}
The resulting wave function depends on quantum number $\zeta$, and we calculate the eigenvalues and eigenfunctions of the equation
\begin{equation}
\mathcal{H}_\chi \Psi_\chi = E \Psi_\chi,
\end{equation}
where $\mathcal{H}_\chi = U_\chi^{-1} \mathcal{H} U_\chi$. The equations for three spinor components of the wave function (when there is no external magnetic field present) are
\begin{align}
&- \frac{1}{2} \frac{d^2}{dr^2} \psi_{\chi,1}(r) + \frac{ \psi_{\chi,1}(r) }{8r^2} + \frac{\zeta^2 \psi_{\chi,1}(r)}{2r^2} - B(r) \psi_{\chi,1}(r)  \nonumber \\
& + V(r) \psi_{\chi,1} + \frac{ \zeta \psi_{\chi,2}(r) }{ \sqrt{2}r^2 } + \frac{ \psi_{\chi,3}(r) }{ 4r^2 } = E \psi_{\chi,1}(r),\\
& - \frac{1}{2} \frac{d^2}{dr^2} \psi_{\chi,2}(r) + \frac{ \zeta \psi_{\chi,1}(r) }{ \sqrt{2}r^2 } + \frac{ 3\psi_{\chi,2}(r) }{ 8r^2 } + \frac{ \zeta^2 \psi_{\chi,2}(r) }{ 2r^2 } \nonumber \\
&+ V(r) \psi_{\chi,2}(r) + \frac{ \zeta \psi_{\chi,3}(r) }{ \sqrt{2}r^2 } = E \psi_{\chi,2}(r) \\
& - \frac{1}{2} \frac{d^2}{dr^2} \psi_{\chi,3}(r) +\frac{ \psi_{\chi,1}(r) }{ 4r^2 } + \frac{ \zeta\psi_{\chi,2}(r) }{ \sqrt{2}r^2 } + \frac{ \psi_{\chi,3}(r) }{ 8r^2 } \nonumber \\
&+ \frac{ \zeta^2\psi_{\chi,3}(r) }{ 2r^2 } + B(r)\psi_{\chi,3}(r) + V(r)\psi_{\chi,3}(r) = E \psi_{\chi,3}(r).
\end{align}
Results for $\zeta=0,\pm 1, \pm 2$ and $\zeta = \pm 3$ are shown in Fig.~\ref{fig:Single_atom_energy_levels zeta}.

We solved the Schr\"odinger equation for a single atom in a lattice cell to find the energy eigenvalues and eigenfunctions  for several of the lowest energy levels.   In our case, the SDOLP has hexagonal symmetry, however, close to the SDOLP minimum at $\vec{r}= (x,y) = (0,0)$ the potential has radial symmetry. Most of the particle areal density is localized near potential minimum, where system is radially symmetric. Using quantum numbers $n$ and $\zeta$ we identify the energy eigenstates.  For a system with radial symmetry there are two good quantum numbers, $n$ and $\zeta$, corresponding to quantum numbers $n_1, n_2$ of a two-dimensional harmonic oscillator as follows $n=n_1+n_2$ and $\zeta=n_1-n_2$, see Table \ref{quantum numbers}.  When $B_\mathrm{\rm{ext}}=0$ the ground state corresponds to $n_1=0,n_2=0$, so $n=0, \zeta = 0$, the first excited state has $n=1, \zeta = \pm 1$. States with $\zeta=0$ are non-degenerate and states with $\zeta=\pm1,\pm2,\dots$ are doubly degenerate at $B_\mathrm{\rm{ext}}=0$. The doubly degenerate levels split in two states for non-vanishing external magnetic field.

\begin{figure}[h]
\centering
\includegraphics[width=0.95\linewidth]{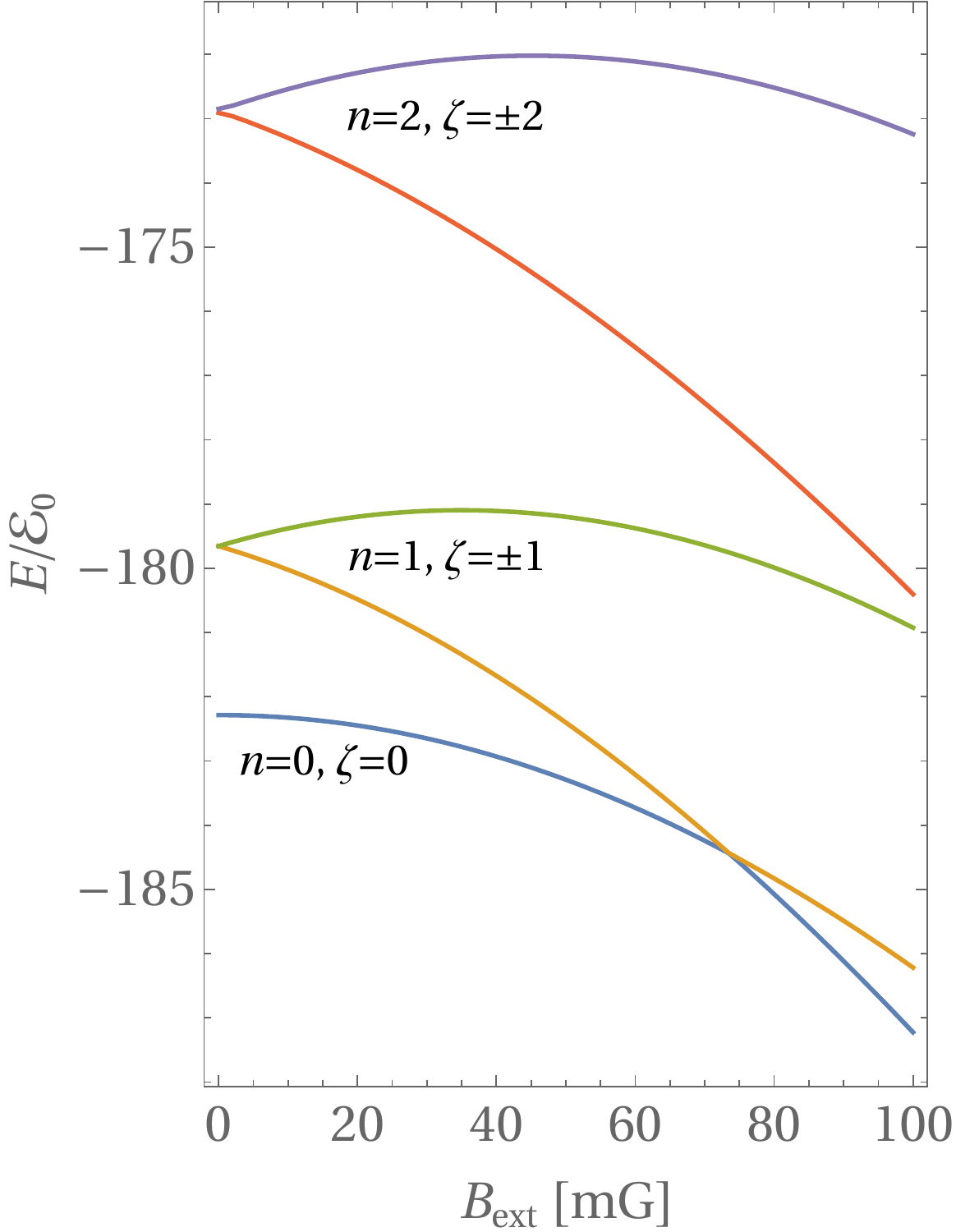}
\caption{The energy eigenvalues of the Schr\"odinger equation as a function of external magnetic field strength.}
\label{fig:Single atom energy levels}
\end{figure}

\begin{figure}[h]
\centering
\includegraphics[width=1\linewidth]{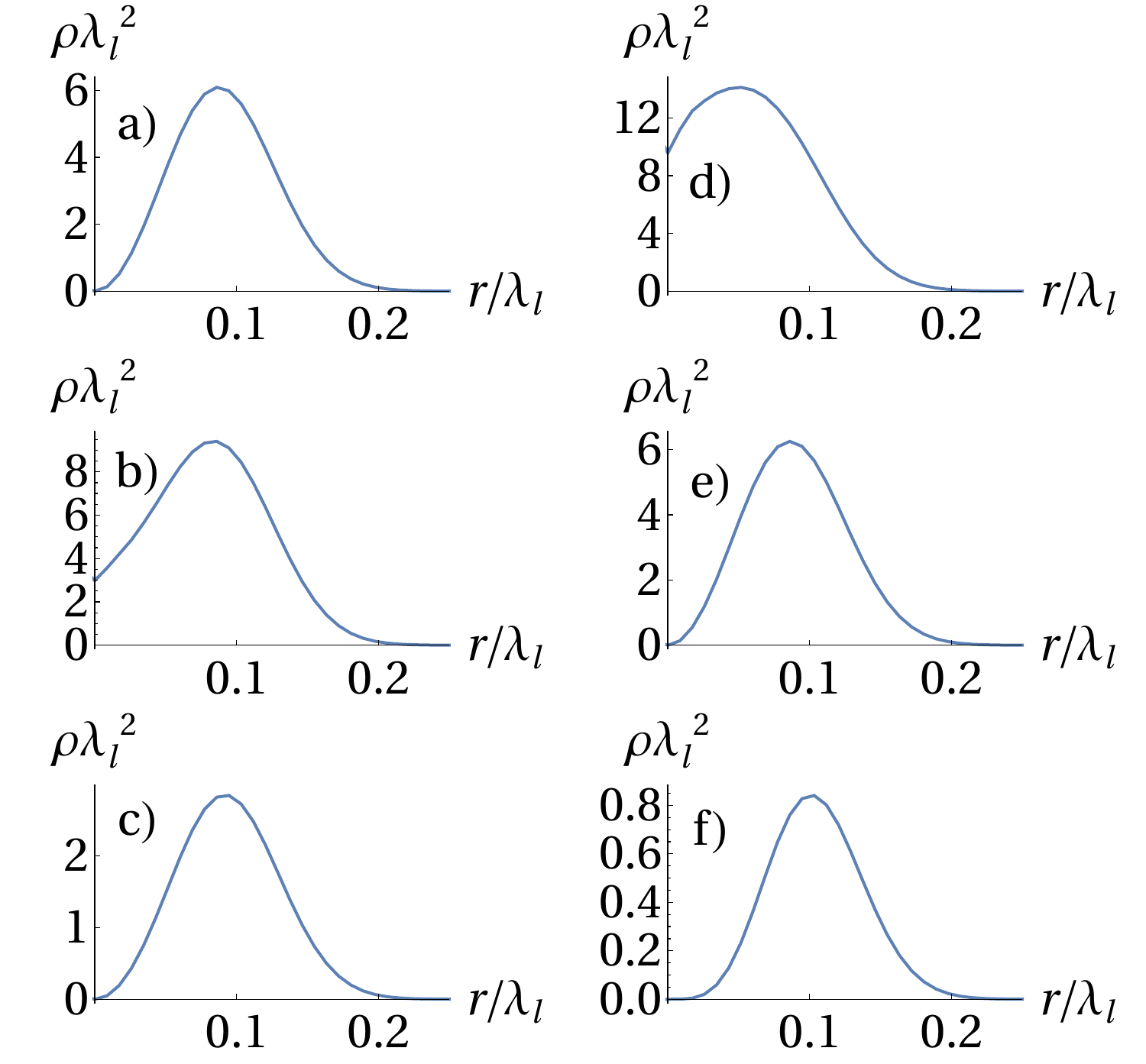}
\caption{Single atom ground state density in the isotropic approximation.  Spinor component (a) $m_F=+1$, (b) $m_F=0$, (c) $m_F=-1$ for $B_\textrm{ext}=40$ mG and (d)  $m_F=+1$ (e) $m_F=0$ and (f) $m_F=-1$ for $B_\textrm{ext}=100$ mG.}
\label{fig:density single atom}
\end{figure}

Figure~\ref{fig:Single atom energy levels} plots the lowest five energy levels as a function of the external magnetic field strength $B_\mathrm{\rm{ext}}$ for the SDOLP laser intensity set equal to 70 W/cm$^2$. Laser wavelength was $\lambda_l = 795.456$ nm and potential had hexagonal symmetry. However, as we can see in Fig.~\ref{fig:density single atom}, density was located close to the center $\vec{r}=(0,0)$, where potential has radial symmetry. For $B_\mathrm{\rm{ext}}=0$, the ground state is non-degenerate but the higher energy levels are doubly degenerate.  The ground state crosses the $n = 1$, $\zeta = -1$ level near $B_\mathrm{\rm{ext}}=73$ mG.

Figure~\ref{fig:density single atom} shows the density of the $m_F=+1, 0$ and $-1$ spinor component of the ground state for $B_\mathrm{\rm{ext}}=40$ mG and $B_\mathrm{\rm{ext}}=100$ mG. Notice that for the lower value of the external magnetic field, the wavefunction in the components $m_F=+1$ and $m_F=-1$ are zero at the origin (in agreement with the total value of $\zeta=0$), and above the transition,(i.e., at $B_\mathrm{\rm{ext}}=100$ mG) the component wavefunctions with $m_F=0$ and $m_F=-1$  go to zero at the origin. The same is true for the case of many atoms in the BEC case. Figure ~\ref{fig:density single atom} should be compared with Fig.~\ref{Density of BEC at 6080mG} in the MT.

\begin{table}
\centering
\resizebox{0.48\textwidth}{!}{%
\begin{tabular}{|c|c|c|c|c|}
\hline
 $n_1 \backslash n_2$ & 0 & 1 & 2 & 3 \\ 
\hline
 0 & $n=0,\zeta=0$ & $n=1,\zeta=-1$ & $n=2,\zeta=-2$ & $n=3,\zeta=-3$ \\ 
 \hline
 1 & $n=1,\zeta=+1$ & $n=2,\zeta=0$ & $n=3,\zeta=-1$ & $n=4,\zeta=-2$ \\ 
 \hline
 2 & $n=2,\zeta=+2$ & $n=3,\zeta=+1$ & $n=4,\zeta=0$ & $n=5,\zeta=-1$ \\ 
 \hline
 3 & $n=3,\zeta=+3$ & $n=4,\zeta=+2$ & $n=5,\zeta=+1$ & $n=6,\zeta=0$ \\ 
 \hline
\end{tabular}
}
\caption{Quantum numbers $n,\zeta$}
\label{quantum numbers}
\end{table}

\begin{figure}
\centering
\includegraphics[width=1\linewidth]{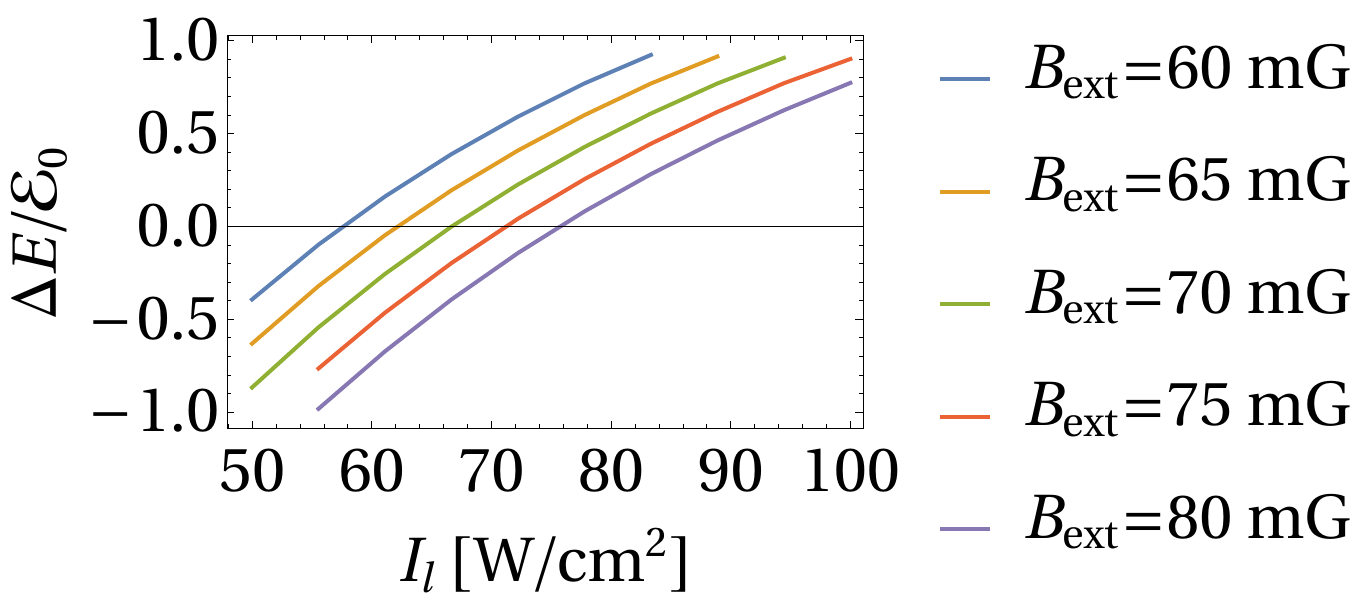}
\caption{Energy difference between $\zeta=0$ and $\zeta=1$ states for fixed value of the external magnetic field as a function of the laser intensity of the beams creating the SDOLP.}
\label{fig:crossing_intensity_dependence}
\end{figure}

Figure \ref{fig:crossing_intensity_dependence} shows the results of numerical calculations of the energy difference between the two lowest energy states in the SDOLP versus laser intensity.   The intensity at the crossing is proportional to the external magnetic field strength. Hence, one could keep external magnetic field constant and sweep across the transition point tuning only the laser intensity.

\end{document}